# Pixel-scale NIR-VIS Spectral Routers Based on 2D Mie-type Metagratings

Yifan Shao[1], Shuhan Guo[1], Rui Chen[1], Yongdi Dang[1], Yi Zhou[1], Yubo Wang[1], Junjie Zhan[1], Jiaqi Yu[1], Bing-Feng Ju[2] and Yungui Ma[1]*

[1]State Key Lab of Modern Optical Instrumentation, Centre for Optical and Electromagnetic Research, College of Optical Science and Engineering; International Research Center (Haining) for Advanced Photonics, Zhejiang University, Hangzhou, 310058, China

[2]The State Key Lab of Fluid Power Transmission and Control, School of Mechanical Engineering, Zhejiang University, Hangzhou, 310027, China

**Corresponding author's E-mail:** yungui@zju.edu.cn

**Abstract**

The out-of-band energy loss caused by in-built color filters significantly degrades the signal-to-noise ratio and the dynamic range of conventional image sensors, which has restricted the attempt to develop ultrahigh-density imaging devices by merely shrinking the pixel size. This issue will be more serious for security cameras which need to collect visible (VIS) light and near-infrared (NIR) photons as well. The existing solutions mostly explore complex photonic nanostructures, which are often too complicated for production. In this work, we demonstrate a pixel-scale spectral router utilizing two-dimensional (2D) $Si_3N_4$ Mie scattering metagratings that can spatially divide NIR (850 nm) and VIS (400-700 nm) light to different pixels at high efficiencies. It has a minimum feature size larger than 360 nm, highly promising for massive production. Compared with the traditional filter design, our router can gain ~ 42% and 30% signal enhancement for NIR and VIS band, respectively. We show that it also has good polarization insensitivity and incident angle tolerance. The NIR-VIS simultaneous imaging is inspected without any complex reconstruction algorithm. Mode analysis indicates that the multipolar scattering of our Mie-type metagratings provides the necessary degrees of freedom to spatially optimize the routing functions for broadband photons.

# 1. Introduction

Spectral engineering is a crucial technique in many vital optical devices and apparatuses, for example, on-chip spectrometers,[1-3] multicolor holograms,[4-6] achromatic metalens,[7, 8] and hyperspectral imagers.[9, 10] The spectral selection technique is extremely significant to color imaging systems in visible and near-infrared regimes, which is usually realized by integrating color filter arrays with photonic sensors.[11-13] However, the energy utilization efficiency of this layout is very limited as each color filter only passes narrow-band photons with the rest totally damped out. This issue will be more serious in the pursuit of ultrahigh-density compact imaging devices due to the degraded features of the signal-to-noise ratio and the dynamic range for smaller pixels.[12-14] Various nanophotonic spectral filters have been developed to improve the transmittance,[15-18] but can hardly resolve the problem as they employed the same mechanism (filtering) to extract color information.

To circumvent the energy loss issue encountered by classic spectral filters, spectral routers (also called color routers) that can spatially divide light to pre-defined detecting sensors according to their wavelengths have been investigated. The color separation gratings were early proposed by Dammann in 1978 using multilevel dielectric structures to separate red (R), green (G) and blue (B) bands to +1, 0 and -1 order directions, respectively.[19] A color separation optical element was realized experimentally based on the similar design by Layet et al. in 1999.[20] However, the large grating period and the long propagation distance from gratings to detectors limit the pixel level integration in imaging application. Recently, plasmonic gratings and multiwavelength nanoantennas have been developed,[21-24] inducing the color routing effect in a subwavelength scale. However, the overall energy efficiency is limited by the intrinsic loss of metals. To solve the issue, all-dielectric color splitters have been proposed, based on the conception of wavelength-dependent phase control such as utilizing nanocolumns or nanostrips.[25, 26] In these designs, the nano-splitters only occupy a small part of one unit cell and there often exists severe crosstalk between different color channels. As a consequence, color reconstruction algorithms through a conversion matrix have to be applied to restore the color information, which is unfavorable in imaging dark scenes.[25-27] In addition, metalens arrays can be employed to route and focus light simultaneously but generally with large pixels as they need sufficient spaces to realize the routing phase profiles.[28-30] The spatial multiplexing method utilized in some works will sacrifice the energy utilization efficiency.[28, 29] Inverse design algorithms have been utilized to design spectral routers with high performance.[31-39] In order to increase degrees of freedom, the device topologies are often optimized with deep subwavelength complicated structures, but in practice hardly implementable using the current state-of-the-art technologies, especially for visible and near-infrared regimes.[32-36] Quite recently, Bayer-type color routers with the feature size of about 100 nm have been demonstrated using the single-layer metasurface.[38, 39] However, their fabrication needs advanced electron beam lithography (EBL) techniques, very costive for industrial production.

In this work, we demonstrate a pixel-scale spectral router based on 2D Mie-type metagratings that is able to route NIR (850 nm) photons and VIS (400-700 nm) light to distinct pixels at high energy efficiencies. Mie scattering has long been exploited to direct light or shape wavefront, but mostly for the far-field radiation.[40-43] Here, the multipole photonic modes of Mie scatterers are fully explored to optimize the distribution of the electromagnetic fields of broadband photons just above the pixel array. Compared with conventional NIR-VIS image sensors built with color filters (having the maximum energy efficiency of 50%), our spectral router can provide much higher spectral routing efficiencies, for example, 71.18% for NIR photons and 64.27% for VIS light. It means that over 35% signal enhancement is obtained. The router with higher efficiencies (82.12% for NIR and 67.46% for VIS, i.e. about 50% signal enhancement) by Mie scatterers with the higher aspect ratio has also been verified in the simulation. The developed spectral router is further inspected on the NIR-VIS imaging function without taking any complex reconstruction algorithm. The meta-structures developed here have relatively large feature sizes above 360 nm, very promising for massive production utilizing the mature technologies such as deep UV lithography[44] or nanoimprinting.[45-48] The key idea to manipulate broadband photons by tailoring mutipolar scattering of big Mie scatterers other than looking for complicated nanostructures as practiced here is believed instrumental for the development of various meta-devices for practical applications.

## 2. Results
### 2.1. Design and Simulation of the Spectral Router

Figure 1a schematically shows the conventional NIR-VIS image sensors consisting of microlenses, color filters and photodetectors, where a large amount of energy is wasted due to the selective absorption of color filters. In this work, we intend to design spectral routers with the function shown in Figure 1b. The spectral router can separate NIR and VIS light onto different pixels to obtain higher spectral routing efficiencies. Then, the routed light can be filtered by color filter arrays to reduce crosstalk between different wavelength channels. As shown in Figure 1c-d, the spectral router proposed here is based on the 2D metagratings. The unit cell of the metagratings has a period of 4 μm × 4 μm, aligned with sixteen pixels on the detecting plane. It means that the size of each pixel is 1 μm × 1 μm, well matching the pixel size of current commercial image sensors. In each unit cell, eight $Si_3N_4$ square nanopillars, which function as Mie scatterers, are arranged on the quartz substrate, covered by SU-8 polymer which have the similar refractive index with $SiO_2$. We use the $Si_3N_4$ nanopillars because $Si_3N_4$ is nearly transparent in the visible and near-infrared bands and compatible with CMOS process technologies.[49] The optical constants of $Si_3N_4$, $SiO_2$ and SU-8 are shown in Figure S1 (Supporting Information). The scattered field $\mathbf{E}_s$ after passing through the metagratings is determined by the interference of electromagnetic multipole modes supported by Mie scatterers, which strongly depend on their geometries and material composition.[50] Therefore, we can optimize the field distributions on the detecting plane of the spectral router at different wavelengths by changing the geometries of $Si_3N_4$ nanopillars. In order to route relatively more B (400-500 nm) light to pixel 1, 4, more G (500-600 nm) light to pixel 5, 8, 9, 12, and more R (600-700 nm) light to pixel 13, 16, nanopillars with three different widths ($w_1$, $w_2$, $w_3$) are taken into consideration.

The optimization process of the spectral router is shown schematically in Figure S3 (Supporting Information). Particle swarm optimization (PSO) algorithm[51, 52] is combined with the finite-difference time-domain numerical simulations (Lumerical FDTD Solutions) for the purpose of optimizing design parameters to obtain high spectral routing efficiency. The spectral routing efficiency at a given wavelength is defined as the ratio of the energy reaching to the corresponding detecting pixels to the total energy illuminated on the unit cell at this wavelength. Formulas to calculate spectral routing efficiency are shown in Section S2 (Supporting Information). The design parameters include widths of $Si_3N_4$ nanopillars ($w_1$, $w_2$, $w_3$), the height of $Si_3N_4$ nanopillars ($h$) and the distance between the detecting plane and the top of $Si_3N_4$ nanopillars ($h_d$). The figure of merit (FOM) is set to be related to spectral routing efficiencies. More details of the optimization could be found in Section S3 (Supporting Information).

After iterative optimization, final parameters are determined: $w_1$ = 360 nm, $w_2$ = 420 nm, $w_3$ = 430 nm, $h$ = 1 μm and $h_d$ = 1.4 μm. Figure 2a-d illustrate the power flow density distributions on the detecting plane (XY plane, Z = 2.4 μm, i.e. 2.4 μm above the substrate) of NIR (850 nm), R (630 nm), G (530 nm) and B (447 nm) light, respectively. It is obvious that NIR light is well focused on the eight pixels aligned with the $Si_3N_4$ nanopillars while VIS light is mainly routed to the eight pixels complementary to NIR light. Figure 2e-h shows the power flow density distributions of the XZ cross section (Y = 0.5 μm), reflecting routing phenomena in the propagation direction at different wavelengths. The spectral routing efficiencies of NIR and VIS channels are shown in Figure 3a. The gray dashed line represents the ideal maximum spectral routing efficiency (50%) of conventional color filters utilized in NIR-VIS image sensors. The area above the dashed line and below the spectral routing efficiency curves is the total enhanced routing energy. The spectral routing efficiency of NIR (850 nm) is 71.18% while the average spectral routing efficiency of VIS band (400-700 nm) is 64.27%, with a peak efficiency of 80.36% at 530 nm.

In order to reduce crosstalk between distinct wavelength channels and extract RGB information, NIR, R, G and B filters can be arranged to corresponding pixels on the detecting plane as shown in Figure S2 (Supporting Information). The spectral routing efficiencies of NIR, R, G and B channels are shown in Figure 3b. The average spectral routing efficiencies of R (600-700 nm), G (500-600 nm) and B (400-500 nm) channels are 57.33%, 75.15% and 63.94%, respectively. G light has the best routing performance in RGB because of the location in the center of VIS band and the relatively large weight set in the optimization. Attributed to $Si_3N_4$ nanopillars with three distinct widths in different regions, spectral routing efficiencies of R and B channels are slightly higher (about 2%) than that of the VIS channel.

**2.2. Electromagnetic Multipolar Scattering Analysis**

In this section, we illustrate the physical process for the broadband light routing from the viewpoint of grating diffraction assisted by the Mie scattering of nanopillar elements. To simplify the theorical analysis, as shown in Figure 3c, the super unit-cell (4μm×4μm) for one-pixel is divided into the group of 1.14μm×1.14μm subcells profiled by the black square which have a 45° rotation angle around the $z$-axis. The subcell is utilized to estimate the diffraction efficiency of the periodical nanopillars and their scattering features as well. Since its size is much larger than the largest operation wavelength, diffraction will happen and dominate the output field pattern, which is the primary reason for the spatial splitting of the colorful photons. Under this situation, the near-field coupling between the nanopillar could also be neglected. Based on these, the scattering features of nanopillar elements are manipulated through topology optimization to modulate the diffraction efficiency of each order in order to generate the desired interference patterns for routing at the detector plane ($z$ = 2.4 μm). Figure 3d plots the diffraction spectral efficiency of different orders for the array of 420nm-width nanopillars calculated by a Fourier transformation for the field at the detector plane. The zero and ±1 diffraction orders are dominant at long wavelengths (>600nm) while high orders are intensified at shorter wavelengths. Controlling the relative amplitudes and phases of these orders affords the degrees of freedom to optimize the interference patterns. Here it is realized by adopting Mie-type $Si_3N_4$ nanopillar scatterers which have strong and wideband light interaction capabilities. To show this, Figure 3e plots the multipolar scattering cross-section spectra of the 420nm-width nanopillar by decomposing the scattered field $\mathbf{E}_s$ into the coherent superposition of various fundamental multipolar modes mainly including electric dipole (ED), magnetic dipole (MD), electric quadrupole (EQ), magnetic quadrupole (MQ) and their summation.[50, 53] Strong resonance could be recognized from the spectra, indicating the existence of strong light-structure interaction. At the specific wavelength of 776 nm, the total scattering cross section is about 1.5 μm$^2$, one third of the area (2 μm$^2$) of the subcell. However, like a multimode fiber, the Mie-scatterers adopted here could support multipolar modes in the wide band. The coexistence and the anti-phase features of the corresponding electric and magnetic modes give rise to very low backscattering (reflection) and relatively smooth transmission spectra as indicated in Figure 3d. In other words, the interference of the multipolar modes will majorly decide the phase of local transmitted field at the end of each nanopillar. Different widths ($w_1$, $w_2$, $w_3$) of nanopillars are used here to optimize the overall efficiencies of R, G, B channels. More details about the multipole decomposition are given in Section S4 (Supporting Information). The electric field distributions in the cross section denoted by the dashed line in the upper panel of Figure 3c at four different wavelengths are shown in Figure Sxx 附件 (Supporting Information). They indicate the spatial interference of Mie scattering field with the incident field can route NIR light to the center of the subcell and VIS light to four corners of the subcell. To further understand the role of Mie scatterers in the routing metagrating, we also draw the distributions of scattered field $\mathbf{E}_s$ and total electric field $\mathbf{E}$ on the detecting plane for one super unit-cell (Supporting Information). The routing effect or desired diffraction patterns for the NIR and VIS light arising from the periodic interference of the optimized Mie scattering field is evidenced.

### 2.3. Experimental Demonstration of Spectral Routing

The metagratings have been fabricated and characterized to experimentally verify the performance of NIR-VIS spectral routing. The metagratings comprised of $Si_3N_4$ square nanopillars are successively manufactured by plasma-enhanced chemical vapor deposition (PECVD), electron beam lithography (EBL), lift-off, reactive-ion etching (RIE) and spin coating SU-8. The fabrication process is shown in Figure S6 (Supporting Information) and Methods. Figure 4b shows the optical microscopy image of the metagratings. Figure 4c shows the top view scanning electron microscopy (SEM) image of the unit cell in the metagratings.

Figure 4d is the tilted view SEM image of the unit cell. As shown in Figure 4a, the homemade microscope system is built to characterize the spectral routing effect. The sample is mounted on a nanopositioner (SmarPOD 110.45.1-SC-HV-NM, SmarACT) to ensure that the detecting plane is exactly on the focal plane of the 100× objective. The spectral router is illuminated by collimated LED light beam with four different colors. Spectra of these light sources are shown in Figure S7 (Supporting Information). The pinhole and the 4$f$ system are used to limit the beam size to 1 mm which is smaller than the sample in order to decrease stray light. Intensity profiles on the detecting plane are magnified and imaged onto the CMOS camera. Measured intensity profiles on the detecting plane at four different central wavelengths are shown in Figure 4e-h, in accordance with the simulation results in Figure 2a-d that produces complementary routing effect between NIR and VIS light. The spectral routing efficiencies of NIR (850 nm), R (630 nm), G (530 nm) and B (447 nm) channels are 65.15%, 60.04%, 66.22% and 60.42%, respectively. The relative efficiencies of these four wavelengths are 71.02%, 67.44%, 72.3% and 66.02%, respectively, which is defined by the ratio of the energy reaching the corresponding detecting pixels to the energy passing through the unit cell at corresponding wavelengths.

**2.4. NIR-VIS Imaging with the Spectral Router**

After verifying the NIR-VIS routing function, we will try the spectral router for NIR-VIS imaging. To illustrate the idea and also because of the technical limitations, we did not integrate the spectral router with color filter arrays. Instead, NIR (850 nm), R (630 nm), G (530 nm), B (447 nm) LEDs are used as the light source separately to mimic the condition where color filter arrays are utilized on the detecting plane. Figure 5a illustrates the experimental setup for NIR-VIS imaging. The object is a structural color sample with 2×3 color blocks having different transmission spectra. The size of each color block is 16 μm × 16 μm. The object is imaged onto the spectral router by a 4$f$ system. As shown in Figure 5b-e, the intensity profiles of each color channel on the detecting plane are measured by a 100× microscope and a CMOS camera. As shown in Figure 5f-i, images of NIR, R, G and B channels are directly reconstructed in terms of the measured spectral routing efficiencies without using a conversion matrix, which is unfavorable in imaging dark scene.[25, 26] As a contrast, original images of NIR, R, G and B channels are obtained by the system without the spectral router (Figure 5j-m). Obviously, reconstructed images are in accordance with original images for all wavelength channels. The color imaging can be realized by combining R, G and B channels (Figure 5n-o), also showing a good accordance.

## 3. Discussion

The measured spectral routing efficiencies are slightly reduced compared with the simulation because of the fabrication errors (e.g. the width, height and the steepness of $Si_3N_4$ pillars) and the interface reflection between SU-8 and free space, which is numerically analyzed in **Section S8 (Supporting Information)**. The spectral routing efficiencies can be further improved by coating anti-reflection films on the substrate and the SU-8 layer. Besides, as shown in Figure 3(d), there exist high-order diffractions, which is not totally captured by our 100× objective (numerical aperture NA = 0.8) for diffraction angles larger than 53°. This leads to the reduction of information on the detecting plane and energy efficiencies in the experiment..

In view of the NA of actual imaging systems, the robustness to oblique incidences is an important characteristic of spectral routers. Simulated spectral responses of the spectral router under different incident angles are shown in Figure S10 (Supporting Information). The spectral router can retain above 60% spectral routing efficiency under incident angles up to 10°, corresponding to the NA of 0.174 in imaging systems. As shown in Figure S11 (Supporting Information), because of the symmetry of $Si_3N_4$ nanopillars in the metagratings, the spectral router is also of polarization-insensitive feature, which is applicable to arbitrary polarization illumination in imaging scenes. Moreover, the spin-coated SU-8 layer can be replaced by $SiO_2$ for industrial production. As shown in Figure S12 (Supporting Information), because SU-8 and $SiO_2$ have similar refractive indexes, the spectral router will retain good performance after replacing SU-8 with $SiO_2$. The solid spacing layer between $Si_3N_4$ nanopillars and the detecting plane will allow the spectral router to be integrated with image sensors, greatly enhancing the mechanical stability. This is in contrast to previously reported spectral routers that separated colors in the free space.[25, 30, 39] The aspect ratios of $Si_3N_4$ nanopillars are less than 3 and the minimum feature size is 360 nm, thus highly promising for massive production, such as based on deep UV lithography[44] and nanoimprint techniques.[45-48] Moreover, as shown in Section S13 (Supporting Information), the spectral routing efficiency can be further improved to 82.12% for NIR photons and 67.46% for VIS light by utilizing $Si_3N_4$ scatterers with the aspect ratio of 6, which means about 50% energy enhancement compared with the traditional filter design.

## 4. Conclusion

In conclusion, we have demonstrated a pixel-scale NIR-VIS spectral router based on 2D $Si_3N_4$ Mie-type metagratings. About 35% signal enhancement (compared with conventional color filter scheme) enabled by the wavelength-dependent routing effect has been applied to NIR-VIS imaging experimentally, without any complex reconstruction algorithm due to the low crosstalk. The CMOS-compatible spectral router supporting massive production provides great potentials for industrial applications, especially high-density NIR-VIS imaging in low-light scenarios. Moreover, our results indicate that all-dielectric Mie scatterers can provide the unique degrees of freedom to manipulate light-matter interactions, which offers a new pathway for the design of novel photonic devices in particular with broadband light features.

## 5. Methods

### 5.1 Numerical Simulation

Numerical simulations in this work are performed by the commercial software Lumerical FDTD Solutions. The optical constants of $Si_3N_4$, $SiO_2$ and SU-8 utilized in the simulation are shown in Figure S1 (Supporting Information). When simulating the 2D metagratings, the 4 μm × 4 μm unit cell with eight $Si_3N_4$ square nanopillars covered in SU-8 is modeled. The periodic boundary condition is employed for X and Y directions while the perfect matched layer condition is employed for Z direction (the propagation direction). To calculate spectral routing efficiencies of the spectral router, sixteen 1 μm × 1 μm frequency domain field and power monitors are aligned with the sixteen detecting pixels respectively and arranged a distance of $h_d$ away from the top of $Si_3N_4$ nanopillars. Particle swarm optimization algorithm is combined with FDTD simulation to design a NIR-VIS spectral router with high spectral routing efficiencies.

### 5.2 Device Fabrication

The flow chart of the fabrication is shown in Figure S6 (Supporting Information): a 1 μm-thick $Si_3N_4$ film is deposited on the 500 μm-thick cleaned quartz substrate by plasma-enhanced chemical vapor deposition (PECVD). A 380 nm-thick ZEP520A positron beam photoresist is then spin-coated. The structural profile of the 2D metagratings is defined by the electron beam lithography (EBL). After the photoresist is developed, a 120 nm-thick Cr layer is evaporated onto the sample by electron beam evaporation. The Cr hard mask is patterned on the $Si_3N_4$ film after a lift-off process. The $Si_3N_4$ nanopillars is fabricated by reactive-ion etching (RIE). Finally, the residual Cr mask is removed and a 2.4 μm-thick SU-8 layer is spin-coated on the sample.

**Supporting Information**

Supporting Information is available from the Wiley Online Library or from the authors.

**Acknowledgements**

The authors are grateful to the partial supports from the NSFC (62075196, 61775195 and 61875174), the NSFC of Zhejiang Province (LXZ22F050001), the National Key Research and Development Program of China (No. 2017YFA0205700) and the Fundamental Research Funds for the Central Universities.

**Data and materials availability**

All data are available in the main text or the supplementary materials.

**Conflict of Interest**

The authors declare no financial conflicts of interest.

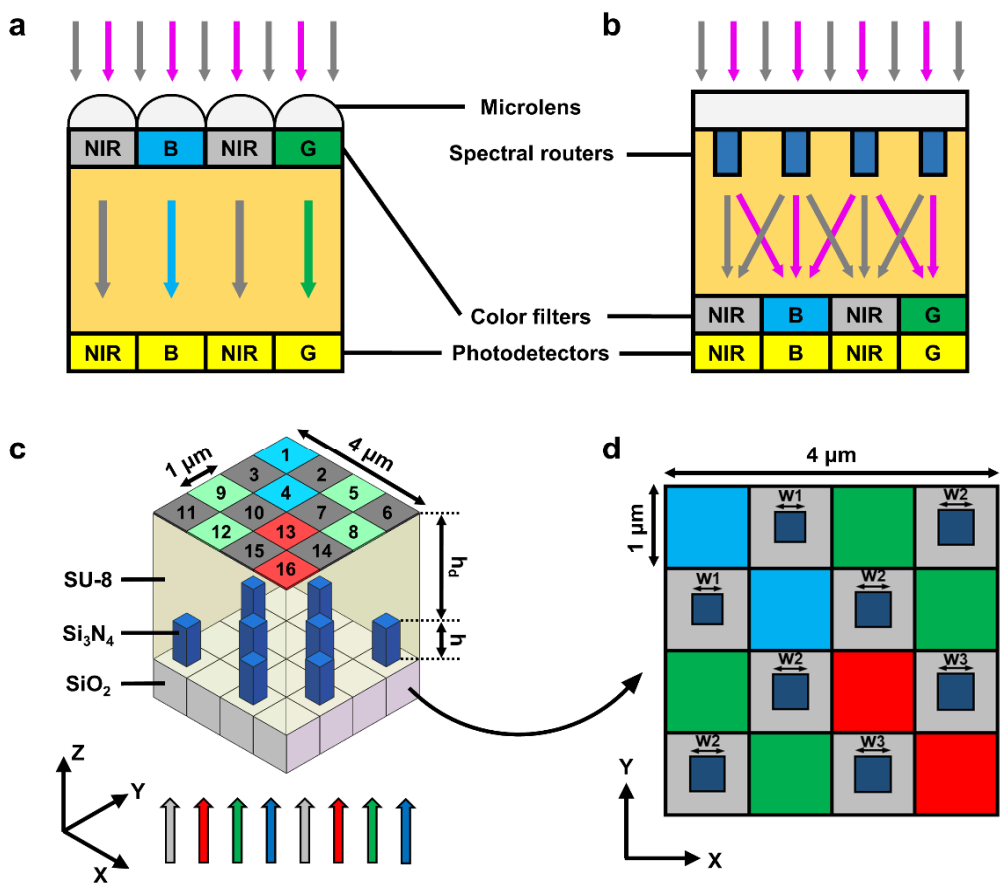

**Figure 1.** a) Schematic side view of the conventional NIR-VIS image sensors with color filters. b) Schematic side view of the NIR-VIS image sensors with spectral routers. Gray arrows and violet arrows in (a-b) represent NIR and VIS light respectively. c) Schematic of the spectral router with $Si_3N_4$ nanopillars covered by SU-8 on a quartz substrate, $h$ is the height of $Si_3N_4$ nanopillars, $h_d$ is the distance between the detecting plane and the top of $Si_3N_4$ nanopillars. d) Schematic top view of the 4 μm × 4 μm unit cell in the spectral router, including sixteen pixels. The eight gray pixels are intended to collect NIR light (850 nm) while the other eight pixels are intended to collect VIS light (400-700 nm). NIR, R, G and B color filters can be correspondingly arranged on the detecting plane to extract color information and eliminate crosstalk between color channels.

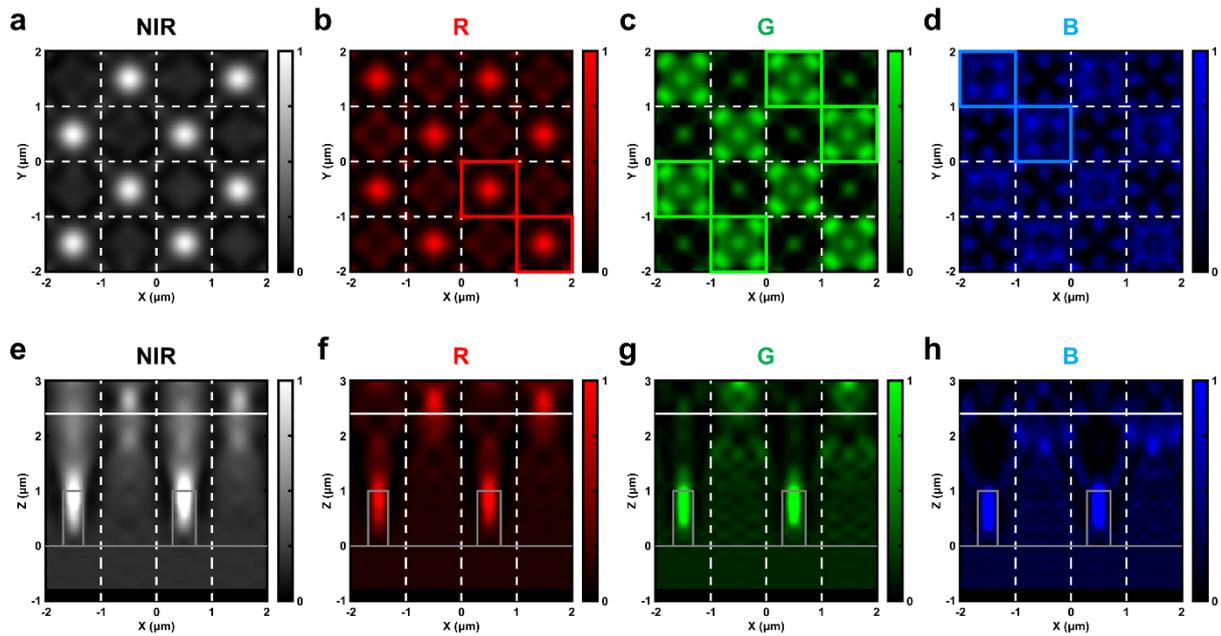

**Figure 2.** a-d) Simulated power flow density distributions on the detecting plane in a unit cell at wavelengths of 850 nm, 630 nm, 530 nm and 447 nm, respectively. The R, G and B boxes in (b-d) represent pixels that can be arranged with R, G and B filters, respectively, while the other eight pixels can be arranged with NIR filters. e-h) Simulated power flow density distributions of the XZ cross section (Y = 0.5 μm) at wavelengths of 850 nm, 630 nm, 530 nm and 447 nm, respectively. The coordinate system is shown in Figure 1c. Gray rectangular boxes represent $Si_3N_4$ nanopillars. The white solid line at Z = 2.4 μm is the detecting plane.

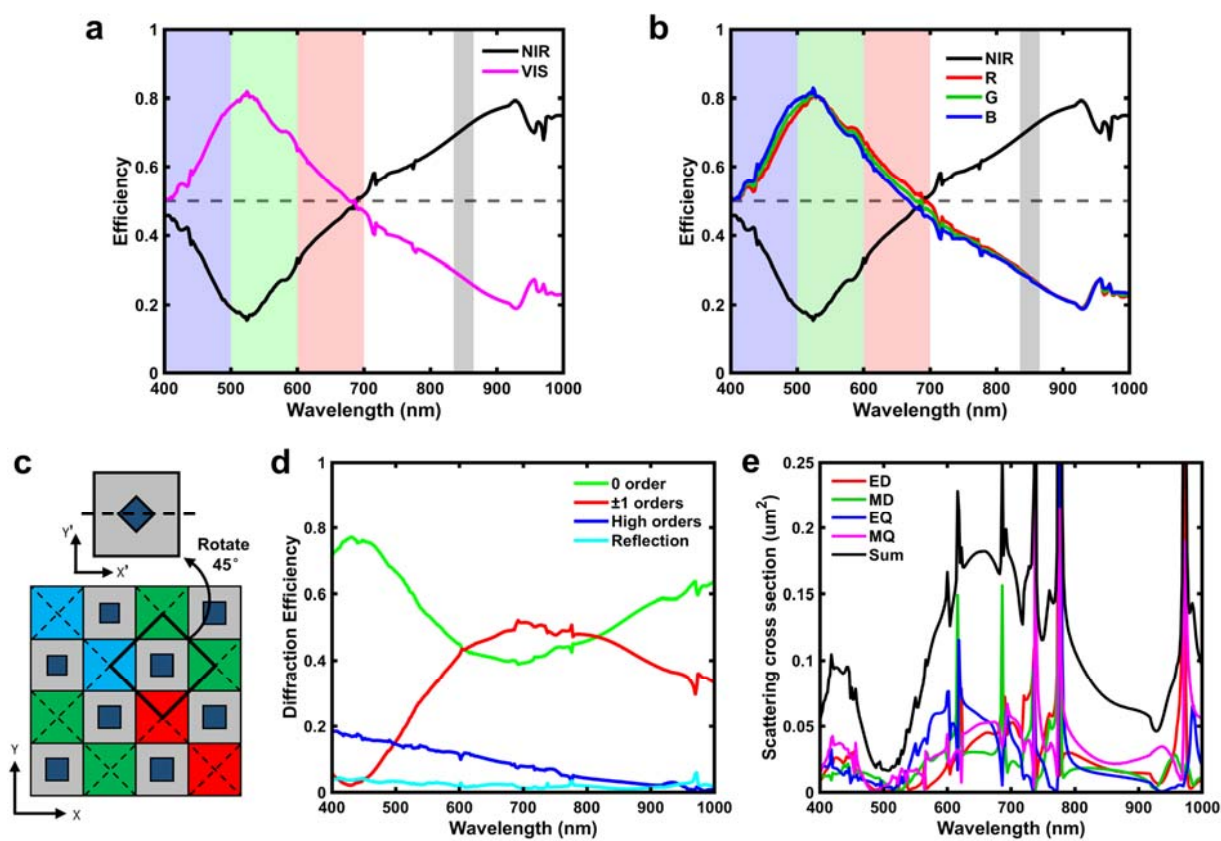

**Figure 3.** a) Simulated spectral routing efficiencies of NIR and VIS channels. b) Simulated spectral routing efficiencies of NIR, R, G and B channels if color filters are arranged. The horizontal dashed line in (a-b) represents the maximum spectral routing efficiency of ideal NIR and VIS filters. The gray, R, G and B strip regions in (a-b) represent the working bands of each color channel respectively. c) The region framed by the black box is regarded as the subcell of the metagrating. The coordinate system is rotated by 45°. d) Diffraction efficiencies of different diffraction orders of the metagrating. The ±1 orders (red line) include the contribution of eight orders, i.e., (0, ±1), (±1, 0), (±1, ±1) orders. (e) Scattering cross sections of the subunit cell with element size of 420 nm for electric dipole (ED), magnetic dipole (MD), electric quadrupole (EQ) and magnetic quadrupole (MQ). The region framed by the black box in (c) is used as the subcell for calculation in (d,e).

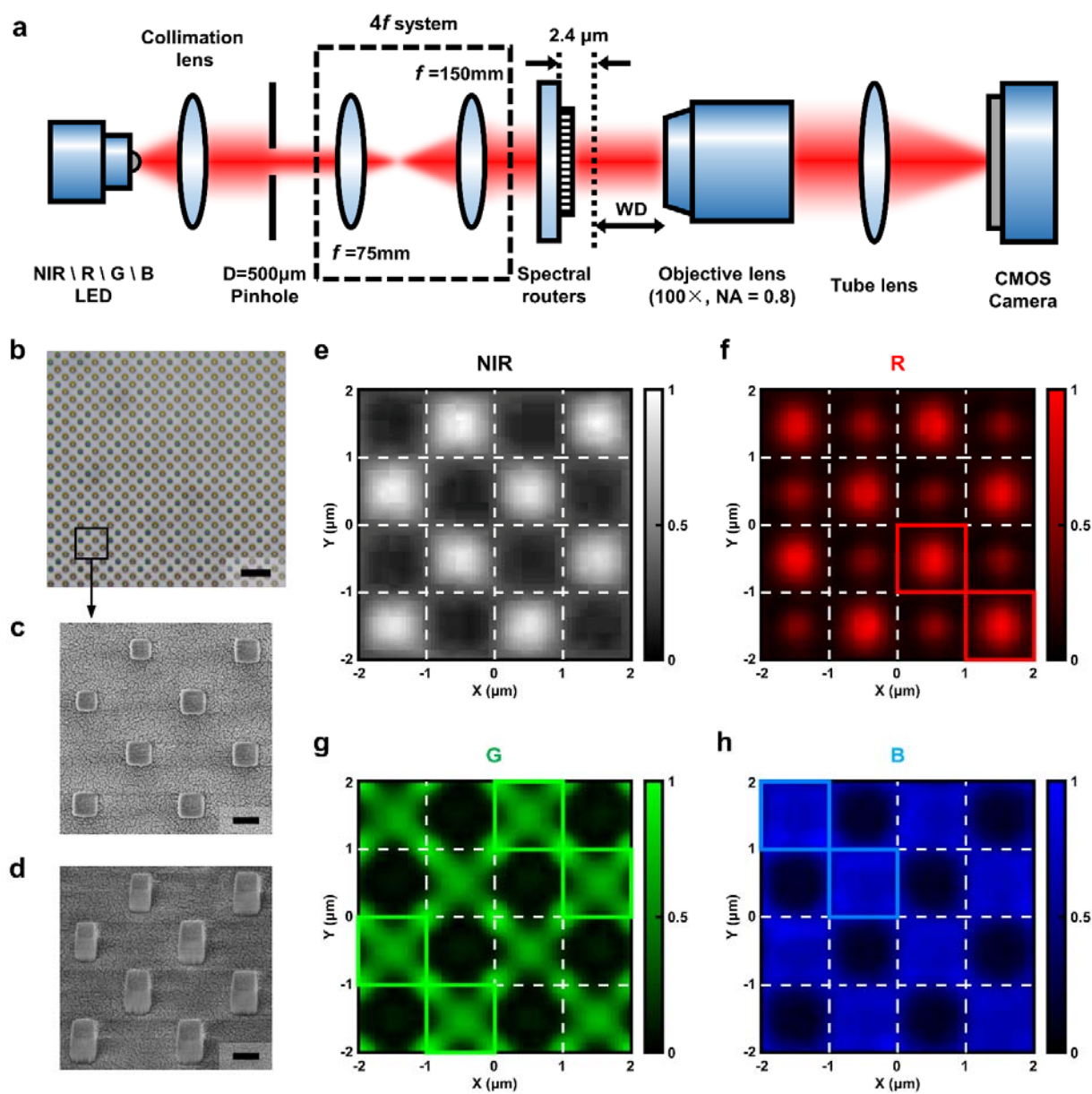

**Figure 4.** Experimental characterizations of the spectral router. a) Schematic of the optical measurement setup. The pinhole and the 4f system are used to limit the beam size. b) Optical microscopy image of the metagratings. Scale bar: 4 μm. c) Top view SEM image of the unit cell in the metagratings. d) Tilted view SEM image of the unit cell in the metagratings. e-h) Measured intensity profiles on the detecting plane in a unit cell at wavelengths of 850 nm, 630 nm, 530 nm and 447 nm, respectively.

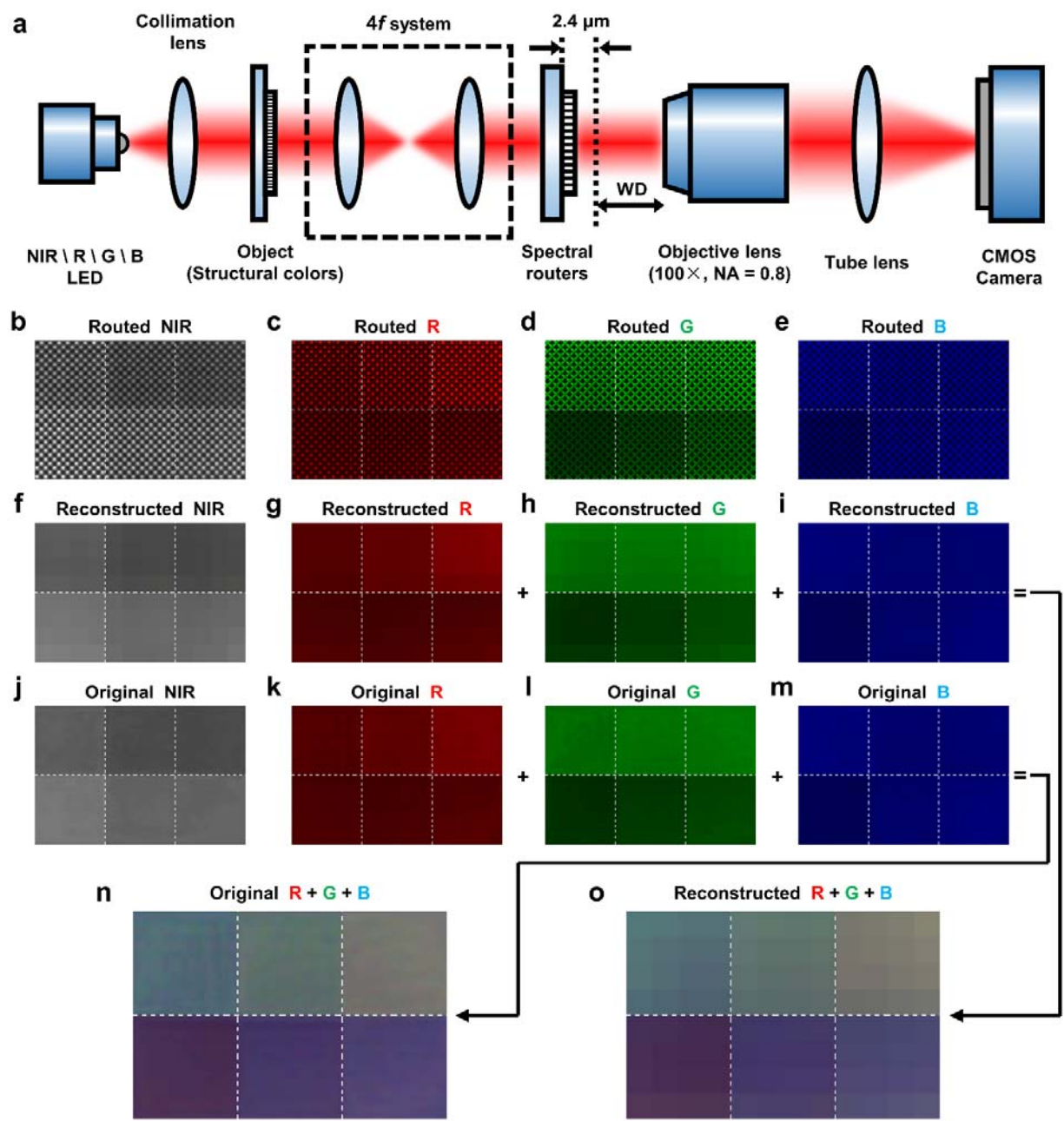

**Figure 5.** NIR-VIS imaging based on the spectral router. a) Schematic of the experimental setup of the NIR-VIS imaging with the spectral router. NIR, R, G and B LEDs are used as the light source one at a time respectively to mimic color filter arrays on the detecting plane. The 4*f* system is used to image the object (2×3 color blocks) onto the spectral router. b-e) Measured intensity profiles of each color channel on the detecting plane after the image is routed. f-i) Reconstructed images of each color channel. j-m) Original images of each color channel obtained by the setup without the spectral router. n) Original color image obtained by (k)+(l)+(m). o) Reconstructed color image obtained by (g)+(h)+(i).

**Table of contents**

The spectral router with the feature size larger than 360nm can spatially separate NIR (850 nm) and VIS (400-700 nm) light to different pixels at high efficiencies, attributed to the optimized multipolar interference supported by $Si_3N_4$ Mie scatterers in the metagrating. It can replace the filter design in traditional NIR-VIS image sensors to improve the signal-to-noise ratio and the dynamic range.

**Pixel-scale NIR-VIS Spectral Routers Based on 2D Mie-type Metagratings**

**Yifan Shao, Shuhan Guo, Rui Chen, Yongdi Dang, Yi Zhou, Yubo Wang, Junjie Zhan, Jiaqi Yu, Bing-Feng Ju and Yungui Ma***

**Figure for TOC**

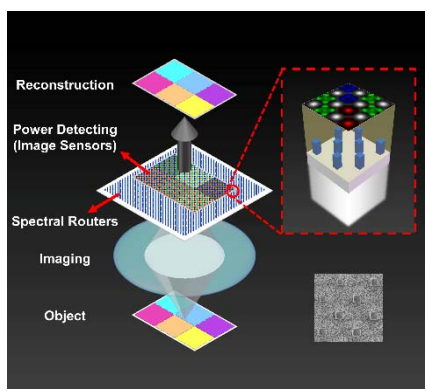

# Supporting Information

**Pixel-scale NIR-VIS Spectral Routers Based on 2D Mie-type Metagratings**

**Yifan Shao[1], Shuhan Guo[1], Rui Chen[1], Yongdi Dang[1], Yi Zhou[1], Yubo Wang[1], Junjie Zhan[1], Jiaqi Yu[1], Bing-Feng Ju[2] and Yungui Ma[1]***

[1]State Key Lab of Modern Optical Instrumentation, Centre for Optical and Electromagnetic Research, College of Optical Science and Engineering; International Research Center (Haining) for Advanced Photonics, Zhejiang University, Hangzhou, 310058, China

[2]The State Key Lab of Fluid Power Transmission and Control, School of Mechanical Engineering, Zhejiang University, Hangzhou, 310027, China

**Corresponding author's E-mail:** yungui@zju.edu.cn

**Section S1. Optical constants of used materials**

The material parameters of $Si_3N_4$, $SiO_2$ and SU-8 are shown in Figure S1. The optical constants of $Si_3N_4$ and SU-8 are measured by an ellipsometer. The refractive index of $SiO_2$ is obtained from the data measured by Malitson in 1965.

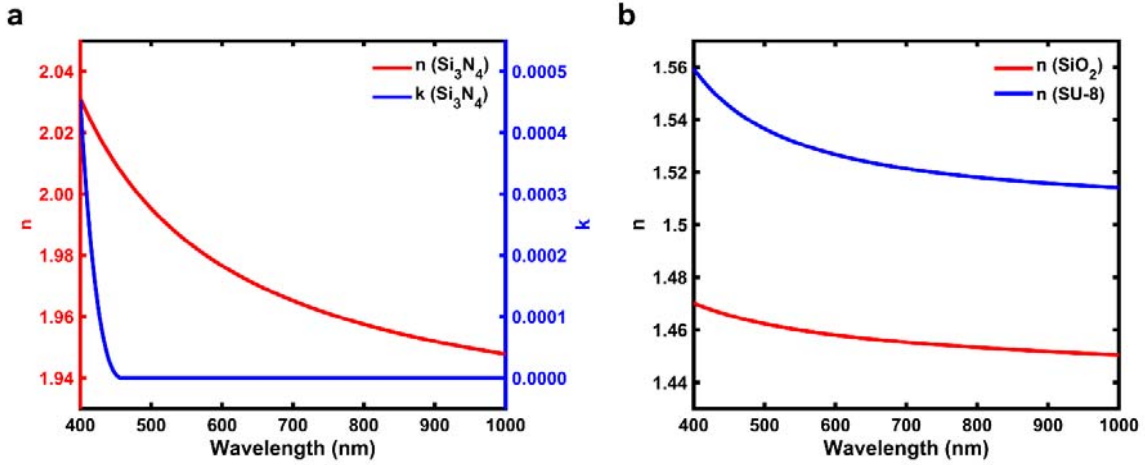

**Figure S1.** (a) The refractive index and the extinction coefficient of $Si_3N_4$. (b) The refractive indexes of $SiO_2$ and SU-8.

**Section S2. Definition of the spectral routing efficiency**

The unit cell including sixteen pixels on the detecting plane is shown in Figure S2. Eight gray pixels (number 2, 3, 6, 7, 10, 11, 14 and 15) are used to accept NIR light. The other eight pixels (number 1, 4, 5, 8, 9, 12, 13 and 16) are used to accept VIS light. The spectral routing efficiency at a certain wavelength is defined as the ratio of the energy reaching to the corresponding detecting pixels to the total energy illuminated on the unit cell at this wavelength.

The spectral routing efficiency of the NIR channel is defined as:

$$T_{NIR} = T_2(\lambda) + T_3(\lambda) + T_6(\lambda) + T_7(\lambda)$$
$$+ T_{10}(\lambda) + T_{11}(\lambda) + T_{14}(\lambda) + T_{15}(\lambda), \quad \lambda = 850 \text{nm} \quad (S1)$$

The spectral routing efficiency of the VIS channel at a given wavelength $\lambda$ is defined as:

$$T_{VIS}(\lambda) = T_1(\lambda) + T_4(\lambda) + T_5(\lambda) + T_8(\lambda) + T_9(\lambda) + T_{12}(\lambda) + T_{13}(\lambda) + T_{16}(\lambda) \quad (S2)$$

where $T_i(\lambda)$ is the ratio of the energy reaching to the pixel (number $i$) to the total energy illuminated on the unit cell at the wavelength $\lambda$.

The average spectral routing efficiency of the VIS band is:

$$T_{VIS} = \frac{1}{\Delta \lambda_{VIS}} \int_{400 \text{ nm}}^{700 \text{ nm}} T_{VIS}(\lambda) d\lambda \quad (S3)$$

where $\Delta\lambda_{VIS} = 300$ nm is the bandwidth of the VIS band.

As shown in Figure S2, in order to extract color information and eliminate crosstalk between color channels, NIR, R, G and B color filters can be arranged to corresponding pixels on the detecting plane. The spectral routing efficiencies of R, G and B channels at a given wavelength $\lambda$ are defined as:

$$T_R(\lambda) = 4 \times [(T_{13}(\lambda) + T_{16}(\lambda)] \tag{S4}$$

$$T_G(\lambda) = 2 \times [T_5(\lambda) + T_8(\lambda) + T_9(\lambda) + T_{12}(\lambda)] \tag{S5}$$

$$T_B(\lambda) = 4 \times [(T_1(\lambda) + T_4(\lambda)] \tag{S6}$$

efficiencies of R, G and B channels are multiplied by the factor of 4, 2 and 4, respectively, for clear comparison with the spectral routing efficiency of the VIS channel.

The average spectral routing efficiencies of R, G and B channels are:

$$T_R = \frac{1}{\Delta\lambda_R} \int_{600\text{ nm}}^{700\text{ nm}} T_R(\lambda)d\lambda \tag{S7}$$

$$T_G = \frac{1}{\Delta\lambda_G} \int_{500\text{ nm}}^{600\text{ nm}} T_G(\lambda)d\lambda \tag{S8}$$

$$T_B = \frac{1}{\Delta\lambda_B} \int_{400\text{ nm}}^{500\text{ nm}} T_B(\lambda)d\lambda \tag{S9}$$

where $\Delta\lambda_R = \Delta\lambda_G = \Delta\lambda_B = 100$ nm are bandwidths of R, G and B bands.

**Figure S2.** Schematic top view of the unit cell with sixteen pixels on the detecting plane.

**Section S3. Optimization process of the spectral router**

Particle swarm optimization (PSO) algorithm is an intelligent stochastic search technique that iteratively tries to improve a candidate solution with regard to a given measure of quality (FOM, i.e. figure of merit).[1, 2] In the optimization of the spectral router, the figure of merit is set as:

$$\text{FOM} = w_{NIR} \times T_{NIR} + w_R \times T_R + w_G \times T_G + w_B \times T_B \qquad (S10)$$

where $T_{NIR}$ is the spectral routing efficiencies of the NIR channel, $T_R$, $T_G$ and $T_B$ are average spectral routing efficiencies of R, G and B channels, respectively. $w_{NIR} = \frac{1}{2}$, $w_R = \frac{1}{8}$, $w_G = \frac{1}{4}$ and $w_B = \frac{1}{8}$ are weights of NIR, R, G and B channels, respectively, which are determined by the area proportion of each color channel in the unit cell. The FOM aims to obtain higher spectral routing efficiencies. The initial design parameters are generated as $w_1 = w_2 = w_3 = 500$ nm (widths of Si$_3$N$_4$ pillars), $h = 800$ nm (the height of Si$_3$N$_4$ pillars) and $h_d = 1.5$ μm (the distance between the detecting plane and the top of Si$_3$N$_4$ pillars). The number of solutions in each generation is set as 10. As shown in Figure S3, in each iteration, the numerical simulation of the device is performed by Lumerical FDTD Solutions so that the FOM can be calculated by the simulated results. The optimization is repeated until the convergence condition (the increase of FOM is less than 0.0001 in 100 generations, i.e. 1000 solutions) is met or the maximum number of generations (500) is reached.

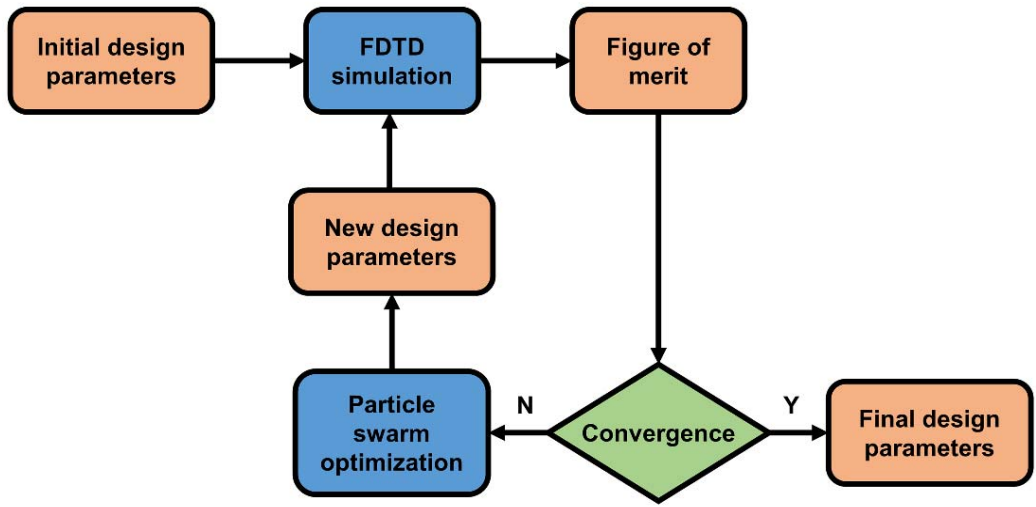

**Figure S3.** Optimization process of the spectral router based on 2D Mie-type metagratings

**Section S4. Multipole decomposition analysis**

Formulas to carry out multipole decomposition are exact expressions which are valid to arbitrarily sized particles of any shape.[3] The formulas to calculate multipole moments are expressed as follows:

$$\mathbf{p}_\alpha = -\frac{1}{i\omega}\left\{\int d^3\mathbf{r}\, \mathbf{J}_\alpha\, j_0(kr) + \frac{k^2}{2}\int d^3\mathbf{r}\,[3(\mathbf{r}\cdot\mathbf{J})r_\alpha - r^2 \mathbf{J}_\alpha]\frac{j_2(kr)}{(kr)^2}\right\} \quad (S11)$$

$$\mathbf{m}_\alpha = \frac{3}{2}\int d^3\mathbf{r}\,(\mathbf{r}\times\mathbf{J})_\alpha \frac{j_1(kr)}{kr} \quad (S12)$$

$$\mathbf{Q}^{\mathrm{e}}_{\alpha\beta} = -\frac{3}{i\omega}\Big\{d^3\mathbf{r}\,[3(r_\beta J_\alpha + r_\alpha J_\beta) - 2(\mathbf{r}\cdot\mathbf{J})\delta_{\alpha\beta}]\frac{j_1(kr)}{kr}$$

$$+ 2k^2\int d^3\mathbf{r}\,[5r_\alpha r_\beta(\mathbf{r}\cdot\mathbf{J}) - (r_\alpha J_\beta + r_\beta J_\alpha)r^2 - r^2(\mathbf{r}\cdot\mathbf{J})\delta_{\alpha\beta}]\frac{j_3(kr)}{(kr)^3}\Big\} \quad (S13)$$

$$\mathbf{Q}^{\mathrm{m}}_{\alpha\beta} = 15\int d^3\mathbf{r}\,\{r_\alpha(\mathbf{r}\times\mathbf{J})_\beta + r_\beta(\mathbf{r}\times\mathbf{J})_\alpha\}\frac{j_2(kr)}{(kr)^2} \quad (S14)$$

where $\mathbf{p}_\alpha$, $\mathbf{m}_\alpha$, $\mathbf{Q}^{\mathrm{e}}_{\alpha\beta}$ and $\mathbf{Q}^{\mathrm{m}}_{\alpha\beta}$ ($\alpha, \beta = x$, y or z) are electric dipole (ED) moment, magnetic dipole (MD) moment, electric quadrupole (EQ) moment and magnetic quadrupole (MQ) moment, respectively. $\omega$ is the angular frequency, $k$ is the wavenumber, $\mathbf{r}$ is the radius vector and $j_i$ are spherical Bessel functions. $\mathbf{J}(\mathbf{r}) = i\omega\epsilon_0[\epsilon_r(\mathbf{r}) - \epsilon_{rs}]\mathbf{E}(\mathbf{r})$ is the scattering electric current density, where $\mathbf{E}(\mathbf{r})$ is the electric field distribution, $\epsilon_0$ is the permittivity of free space, $\epsilon_r(\mathbf{r})$ is the relative permittivity distribution and $\epsilon_{rs}$ is the relative permittivity of the surrounding medium. In addition, there is no need to introduce toroidal multipole moments separately, which are the high-order terms in the electric multipole moments.

The total scattering cross section, i.e. the sum of the contributions from different multipole moments, is expressed as:

$$C^{\mathrm{total}}_{sca} = C^{\mathrm{p}}_{sca} + C^{\mathrm{m}}_{sca} + C^{\mathrm{Q^e}}_{sca} + C^{\mathrm{Q^m}}_{sca} + \cdots$$

$$= \frac{k^4}{6\pi\epsilon_0^2}\left[\sum_\alpha\left(|\mathbf{p}_\alpha|^2 + \left|\frac{\mathbf{m}_\alpha}{c}\right|^2\right) + \frac{1}{120}\sum_{\alpha\beta}\left(|k\mathbf{Q}^{\mathrm{e}}_{\alpha\beta}|^2 + \left|\frac{k\mathbf{Q}^{\mathrm{m}}_{\alpha\beta}}{c}\right|^2\right) + \cdots\right] \quad (S15)$$

where $C^{\mathrm{p}}_{sca}$, $C^{\mathrm{m}}_{sca}$, $C^{\mathrm{Q^e}}_{sca}$ and $C^{\mathrm{Q^m}}_{sca}$ are scattering cross sections of ED, MD, EQ and MQ, respectively.

As shown in Figure 3c of the main text, the region framed by the black box is used to perform the multipole decomposition to show the existences of localized electromagnetic modes in $Si_3N_4$ pillars. The region is rotated by 45° in the Cartesian coordinate system. Electric field distributions of the diagonal cross section at wavelengths of 850 nm, 630 nm, 530 nm and 447 nm are shown in Figure 3c of the main text. Scattering cross sections of different multipole modes are shown in Figure S4 (a-b), which describe the capacity of removing energy from the incident plane wave into the scattered field. Because localized electromagnetic modes at most wavelengths in the working band are not very strong, the spectral routing efficiency spectrum at these wavebands is smooth, while some resonant points shown in the scattering cross section spectrum supported by relatively strong multipole modes will cause disturbances of the spectral routing efficiency spectrum. For example, the resonance at the wavelength of 776nm is mainly dominated by magnetic dipole (MD) mode and electric quadrupole (EQ) mode. The localized electric and magnetic fields at the wavelength of 776 nm are illustrated in Figure S4 (c-d). Besides, some peaks in the spectral routing efficiency spectrum are faint because the total electric field distributions $\mathbf{E}(\mathbf{r})$ are the coherent superposition of incident field $\mathbf{E}_i(\mathbf{r})$ and scattered field $\mathbf{E}_s(\mathbf{r})$ which is determined by the interference of electromagnetic multipole modes supported by $Si_3N_4$ Mie scatterers.

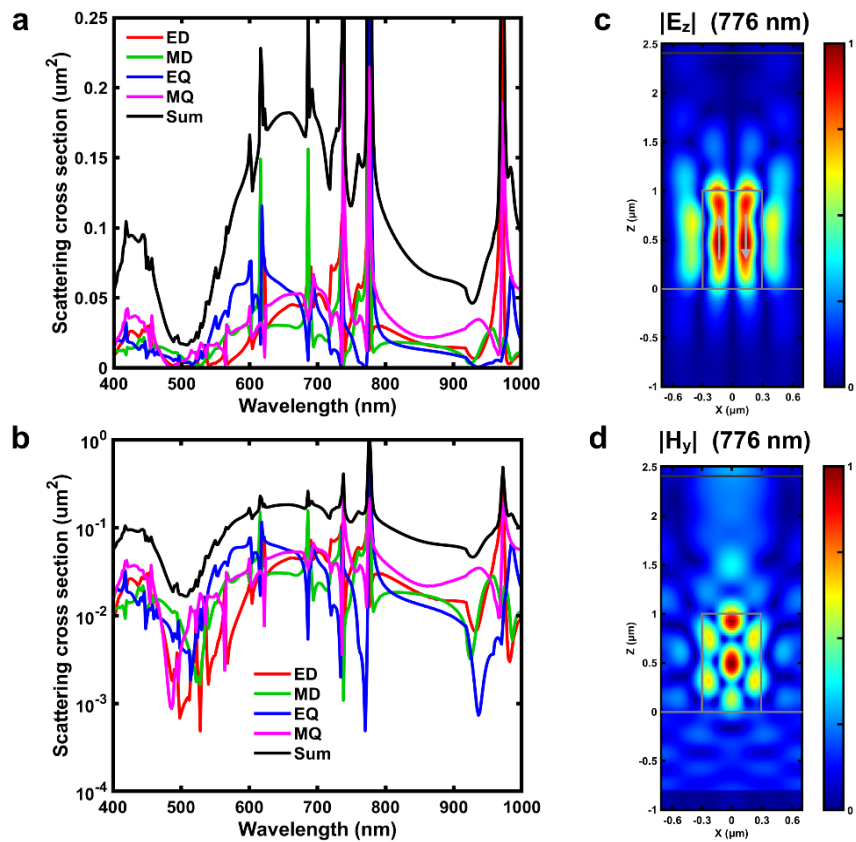

**Figure S4.** Scattering cross sections plotted in (a) linear coordinate and (b) logarithmic coordinate in terms of electric dipole (ED), magnetic dipole (MD), electric quadrupole (EQ) and magnetic quadrupole (MQ). Magnitudes of (a) $E_x$ and (b) $H_y$ of the diagonal cross section. Gray arrows in (c) represent directions of localized electric field.

**Section S5. Scattered electric field and total electric field**

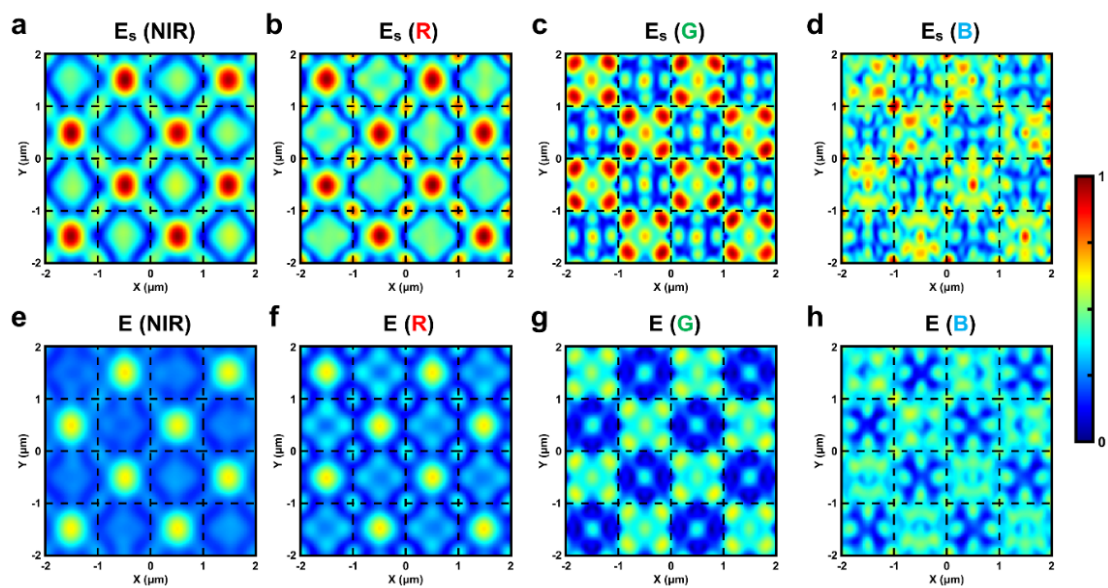

**Figure S5.** (a-d) Scattered electric field **E**$_s$ distributions and (e-h) total electric field **E** distributions on the detecting plane in a unit cell at wavelengths of 850 nm, 630 nm, 530 nm and 447 nm, respectively.

**Section S6. Fabrication of the spectral router**

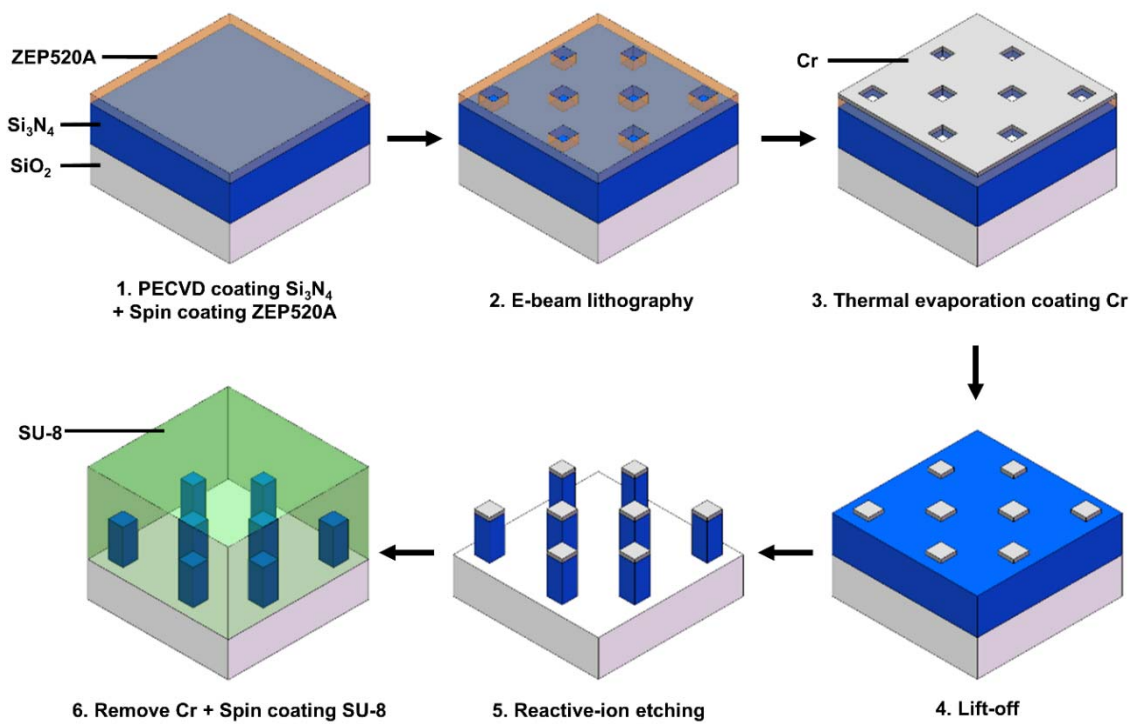

**Figure S6.** Flow chart of the fabrication of the spectral router

**Section S7. Spectra of the light sources**

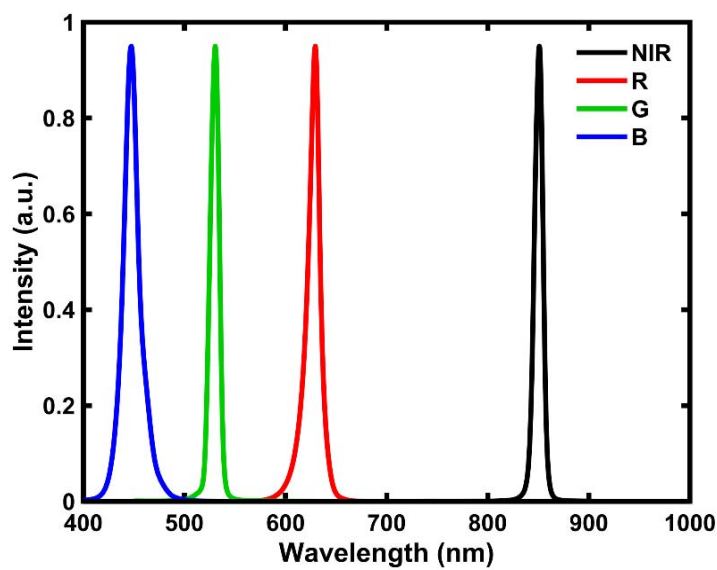

**Figure S7.** Spectra of the NIR, R, G and B LED light sources used in the experimental setup

**Section S8. The actual condition of the spectral router**

Schematic side view of the actual condition of the spectral router is illustrated in Figure S8(a). The thickness of SU-8 is 2.4 μm rather than infinitely thick in the simulation, thus the detecting plane is on the interface between SU-8 and air actually. As shown in Figure S8 (b-c), both spectral routing efficiencies of NIR and VIS channels are slightly reduced when considering the interface while the relative spectral routing efficiencies (i.e. normalized by the transmittance) are similar. It means that the reduced efficiencies are mainly caused by the reflection of the interface, which can be improved by coating anti-reflection films on the substrate and SU-8 to improve the transmittance of the device. The interfacial reflection will also lead to the interference effect, which can be observed by the slight disturbances of the routing spectrum. Figure S8 (d-g) illustrate simulated power flow density distributions on the detecting plane while Figure S8 (h-k) illustrate simulated power flow density distributions of the XZ cross section of NIR (850 nm), R (630 nm), G (530 nm) and B (447 nm) light, respectively, when considering the interface.

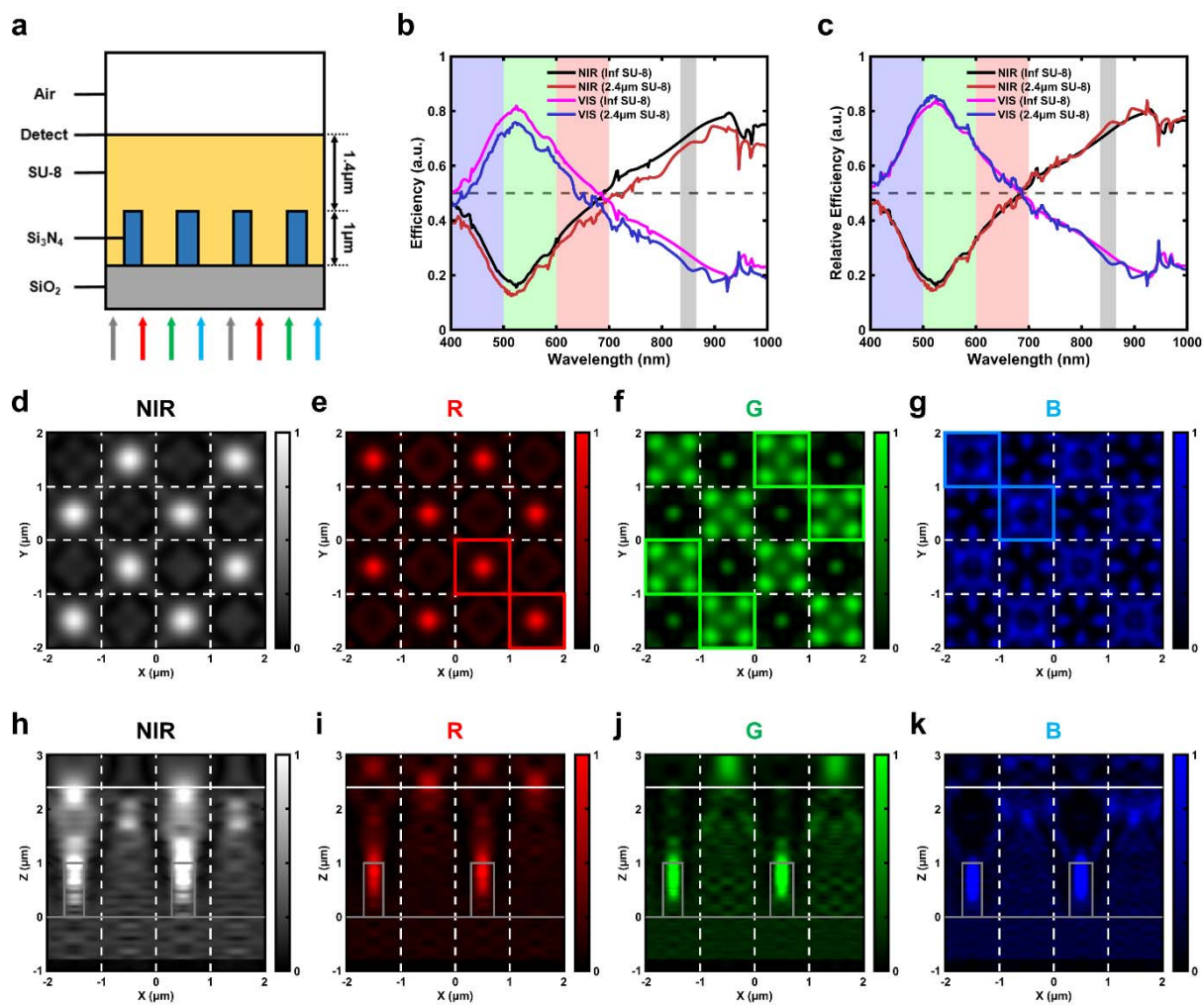

**Figure S8.** Simulation to mimic the actual condition of the spectral router. (a) Schematic side view of the actual condition of the spectral router. The detecting plane is on the interface between SU-8 and air. (b) Comparison between the spectral routing efficiencies when the thickness of SU-8 is infinite and 2.4 μm. (c) Comparison between the relative spectral routing efficiencies when the thickness of SU-8 is infinite and 2.4 μm. (d-g) Simulated power flow density distributions on the detecting plane in a unit cell at wavelengths of 850 nm, 630 nm, 530 nm and 447 nm, respectively, when considering the interface. (h-k) Simulated power flow density distributions of the XZ cross section (Y = 0.5 μm) at wavelengths of 850 nm, 630 nm, 530 nm and 447 nm, respectively, when considering the interface.

## Section S9. Diffraction efficiencies of different diffraction orders

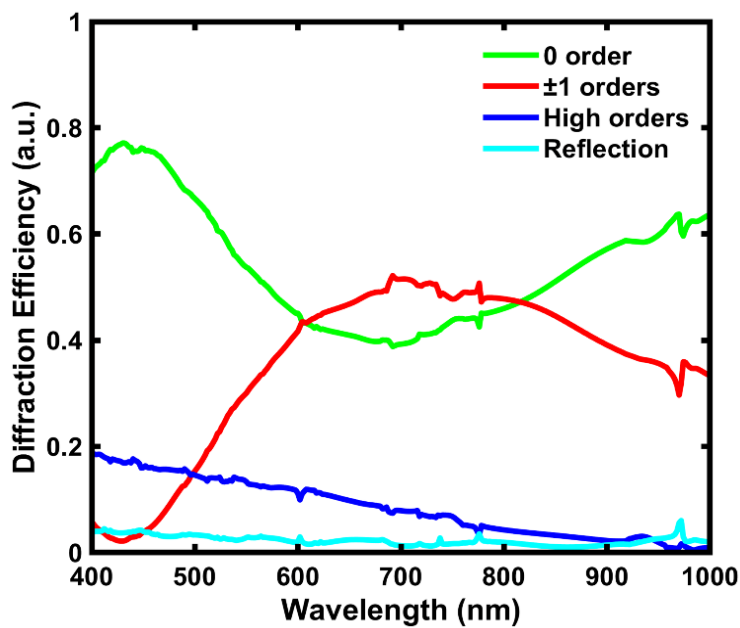

**Figure S9.** Diffraction efficiencies of different diffraction orders of the metagrating. The ±1 orders (red line) contains eight orders, including (0, ±1), (±1, 0), (±1, ±1) orders. The region framed by the black box in Figure 3c of the main text is used as the subcell to calculate the diffraction efficiencies.

## Section S10. Tolerance to incident angles of the spectral router

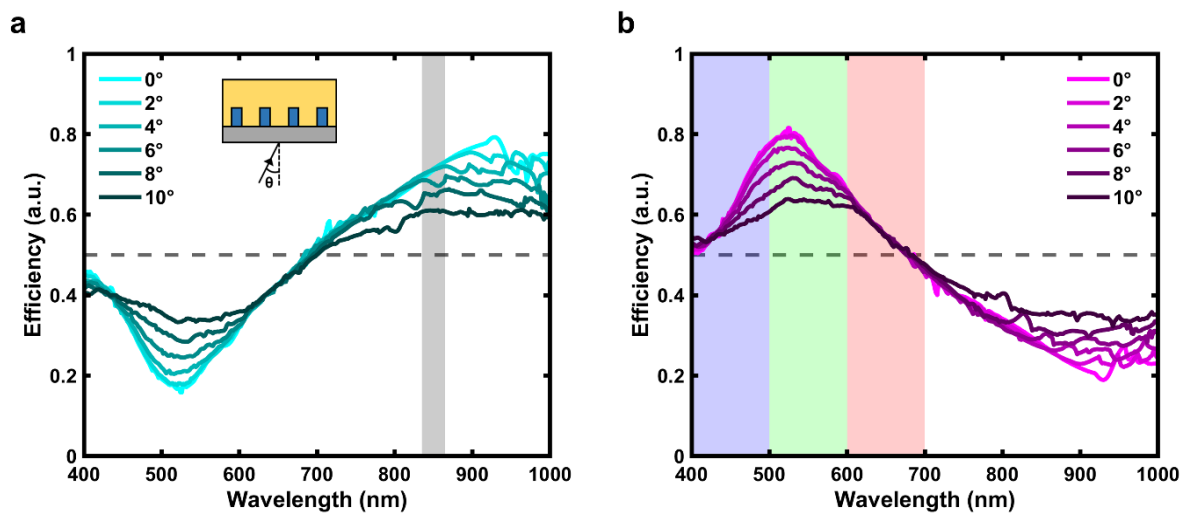

**Figure S10.** Performance dependence on incident angles. Simulated spectral routing efficiencies of (a) NIR and (b) VIS channels under different incident angles. The inset in (a) shows the definition of the incident angle $\theta$.

**Section S11. Polarization insensitivity of the spectral router**

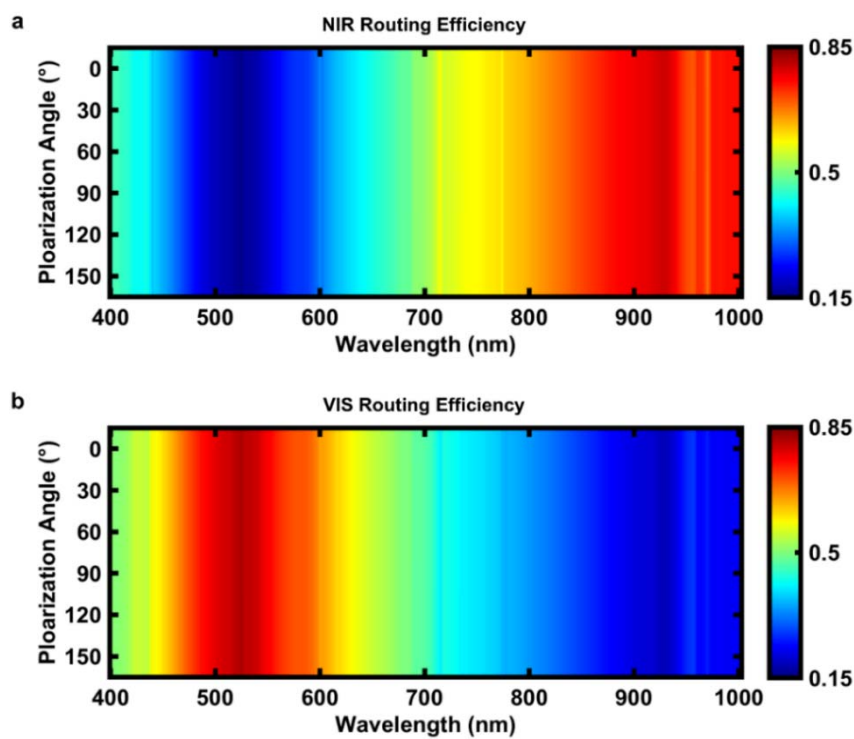

**Figure S11.** Performance dependence on the polarization angle. Simulated spectral routing efficiencies of (a) NIR and (b) VIS channels with different polarization angles.

**Section S12. The device performance when replacing SU-8 with SiO$_2$**

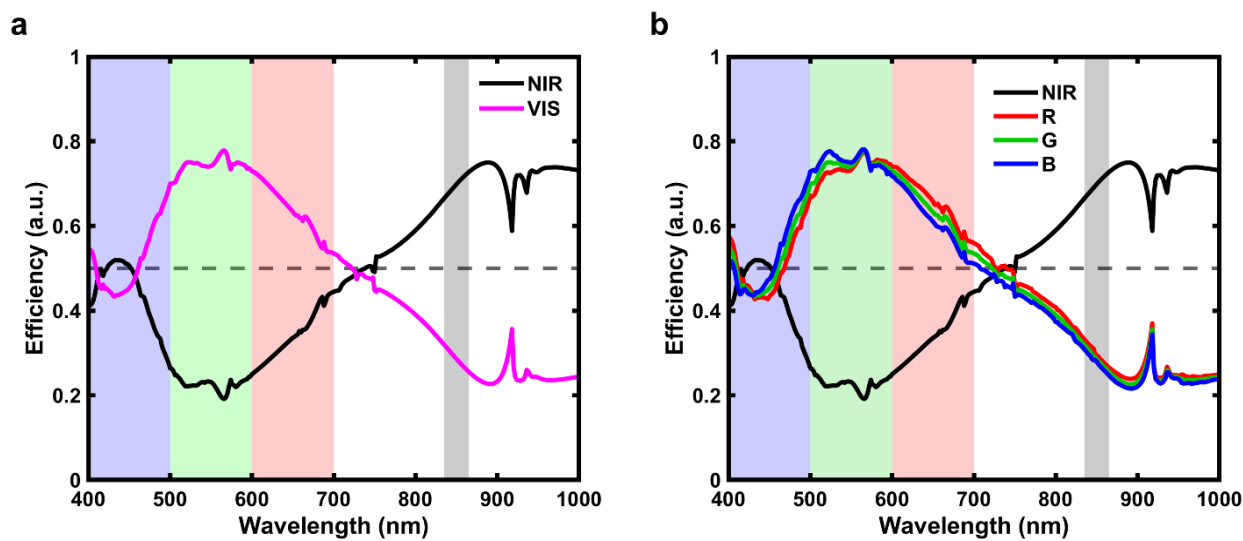

**Figure S12.** (a) Simulated spectral routing efficiencies of NIR and VIS channels after replacing SU-8 with SiO$_2$ without changing geometry parameters of Si$_3$N$_4$ pillars. (b) Simulated spectral routing efficiencies of NIR, R, G and B channels after replacing SU-8 with SiO$_2$ without changing geometry parameters of Si$_3$N$_4$ pillars.

**Section S13. The spectral router with the signal enhancement of about 50%**

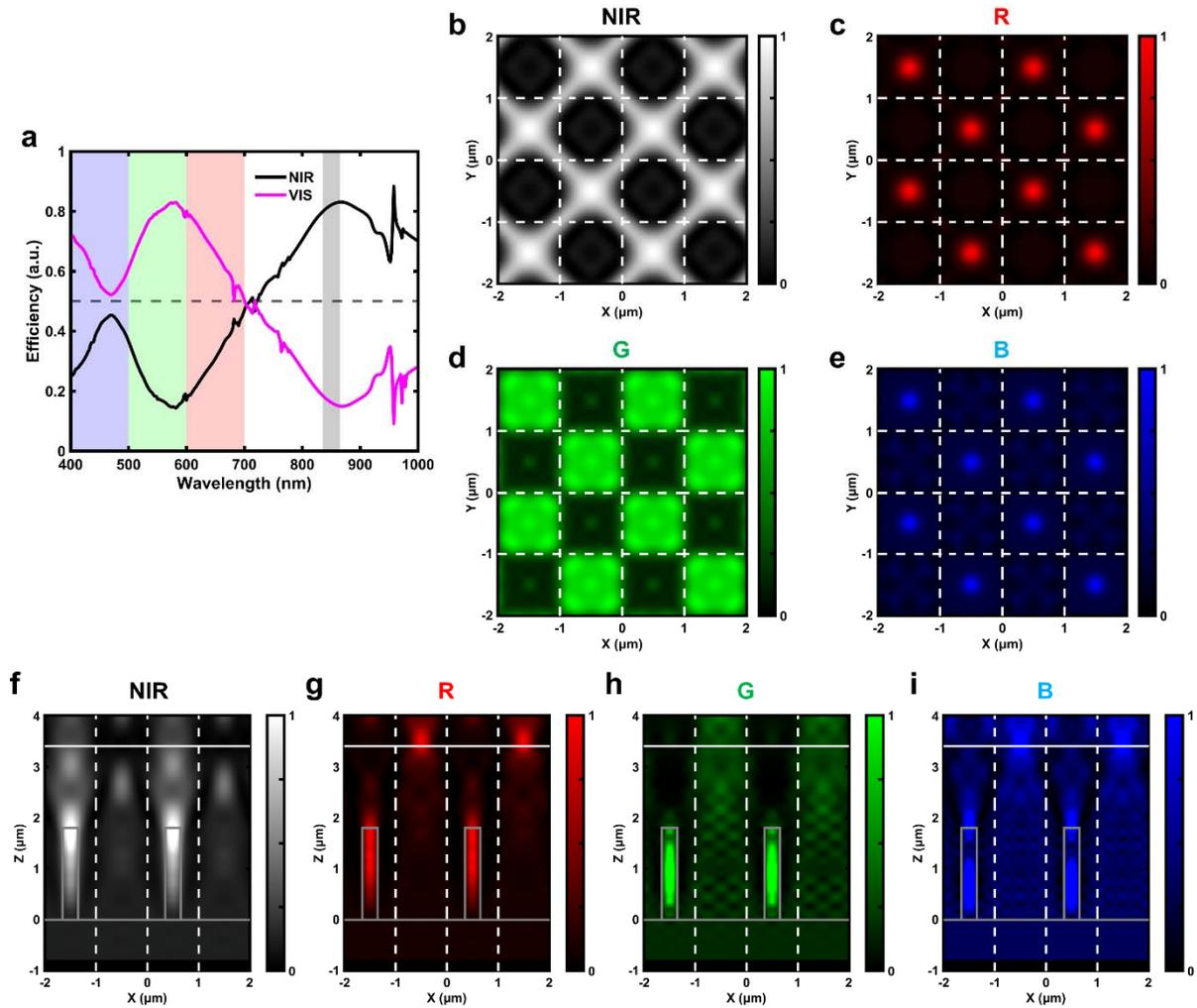

**Figure S13.** Simulation results of the spectral router with higher efficiencies. (a) Simulated spectral routing efficiencies of NIR and VIS channels. The horizontal dashed line represents the maximum spectral routing efficiency of ideal NIR and VIS filters. (b-e) Simulated power flow density distributions on the detecting plane in a unit cell at wavelengths of 850 nm, 630 nm, 530 nm and 447 nm, respectively. (f-i) Simulated power flow density distributions of the XZ cross section (Y = 0.5 μm) at wavelengths of 850 nm, 630 nm, 530 nm and 447 nm, respectively. The white solid line at Z = 3.4 μm is the detecting plane.

In this work, in order to reduce the processing difficulty, the minimum feature size and the height of the $Si_3N_4$ scatterers are designed as 360 nm and 1 μm, respectively. Here, we have also designed a spectral router with higher efficiencies by utilizing Mie scatterers with the higher aspect ratio. The width of $Si_3N_4$ pillars is $w_1 = w_2 = w_3 = 300$ nm. The height of $Si_3N_4$ pillars is $h = 1.8$ μm. The distance between the detecting plane and the top of pillars is $h_d = 1.6$ μm. Figure S13 (a) shows simulated spectral routing efficiencies of NIR and VIS channels. The efficiency of the NIR channel is 82.12%, and the average efficiency of the VIS band is 67.46%. It means that about 50% signal enhancement can be obtained compared with the conventional color filter scheme. The simulated power flow density distributions of the detecting plane (XY plane) and the XZ cross section (Y = 0.5 μm) at different wavelength channels are illustrated in Figure S13 (b-i), respectively, which show the better spectral routing effect between NIR and VIS channels.